# Strong single-ion anisotropy and anisotropic interactions of magnetic adatoms induced by topological surface states

Z. L. Li,[1] J. H. Yang,[1] G. H. Chen,[1] M.-H. Whangbo,[2] H. J. Xiang,[1,*] X. G. Gong[1,†]

[1]*Key Laboratory of Computational Physical Sciences (Ministry of Education), State Key Laboratory of Surface Physics, and Department of Physics, Fudan University, Shanghai 200433, People's Republic of China*

[2]*Department of Chemistry, North Carolina State University, Raleigh, North Carolina 27695-8204, USA*

**Abstract**

The nature of the magnetism brought about by Fe adatoms on the surface of the topological insulator $Bi_2Se_3$ was examined in terms of density functional calculations. The Fe adatoms exhibit strong easy-axis magnetic anisotropy in the dilute adsorption limit due to the topological surface states (TSS). The spin exchange J between the Fe adatoms follows a Ruderman-Kittel-Kasuya-Yosida (RKKY) behavior with substantial anisotropy, and the Dzyaloshinskii-Moriya (DM) interaction between them is quite strong with $|D/J| \approx 0.3$ under the mediation by the TSS, and can be further raised to ~0.6 by an external electric field. The apparent single-ion anisotropy of a Fe adatom is indispensable in determining the spin orientation.

*Introduction.* In a topological insulator (TI) the bulk state has a band gap but the topological surface state (TSS) is gapless due to the protection by the time reversal symmetry (TRS). In three-dimensional (3D) TIs, $Bi_2Se_3$, $Bi_2Te_3$ and $Sb_2Te_3$ [1], the bulk band gap is nontrivial, and the TSS is robust with a single Dirac cone at $\Gamma$ point [2-4] that originates from spin-orbit coupling (SOC) [5]. The TRS of a TI can be broken by introducing magnetic atoms on its surface [6-12], leading to a band gap opening at the Dirac point so that TIs with their surface doped with magnetic atoms can provide useful spintronics applications [6,11]. The interactions between magnetic adatoms on a TI surface are known to follow a RKKY behavior [8,9,11]. In describing these interactions using a model $\bm{k}\cdot\bm{p}$ Hamiltonian, both the anisotropic (Heisenberg-Ising) spin exchange and the antisymmetric (i.e., DM) exchange are taken into consideration [11]. As for the preferred spin-orientation [i.e., the single-ion anisotropy (SIA)] of magnetic adatoms, which opens a gap at the Dirac point when their spins are parallel to the c-axis [3], there exists controversy [3,7]. In accounting for the magnetism mediated by the TSS, therefore, it is necessary to investigate it on the basis of first-principles methods without neglecting the SIA.

In this Letter, we investigate the nature of the magnetism arising from magnetic adatoms on a TI surface by performing density functional calculations for $Bi_2Se_3$ with magnetic Fe atoms adsorbed on its (111) surface and by analyzing results of the calculations in terms of the energy-mapping method [13]. We show that, in the dilute adsorption limit, a Fe adatom exhibits a strong easy-axis SIA due to the spin polarized Bi and Se atoms surrounding it, the spin exchange J between the magnetic adatoms is strongly anisotropic and follows a RKKY-like behavior, and the DM interaction between them is strong with $|D/J| \approx 0.3$ due to the mediation by the TSS and can be enhanced further by external electric field, indicating potential applications of TIs with magnetic adatoms on their surface to spintronics.

*Calculations and structures.* Our density functional calculations were performed using the Vienna *ab initio* simulation package [14] with the projector-augmented-wave method [15,16], the generalized gradient approximation by Perdew, Burke, and Ernzerhof [17] for exchange-correlation functional, and a plane-wave cut-off energy of 340 eV. Experimentally, $Bi_2Se_3$ with a few quintuple layers (QLs) or even only one QL were prepared to study the TSS [2,18]. In this Letter, we simulate the (111) surface of $Bi_2Se_3$ by using a slab of one quintuple layer (1QL-slab) of $Bi_2Se_3$ with the vacuum layer more than 10 Å. The structure of 1QL-slab with Fe atoms adsorbed on one surface was fully relaxed using one k-point (i.e., $\Gamma$). With the resulting optimized structure, 4×4×1 supercell calculations were carried out using a set of 3×3×1 k-points for one Fe adsorption per supercell as well as for cases of two Fe adatoms per supercell when the distance R between them is short. However, 8×4×1 supercell calculations were performed using a set of 2×3×1 k-points for cases of two Fe adatoms per supercell if their distance R is large (see below). The use of the large supercells is necessary to simulate the dilute adsorption limit. All calculations were performed including SOC on all atoms unless otherwise stated. In our calculations the effect of an external electric field along the c-direction was included by using the planar dipole layer method [19]. We also simulated the Fe-adsorbed (111) surface of $Bi_2Se_3$ by using a slab of two QLs (2QL-slab) and that of four QLs (4QL-slab), to find that the use of the 1QL-slab model for the (111) surface of $Bi_2Se_3$ is sufficient for discussing the problems under consideration [20].

The top views of the 4×4 and 8×4 supercells of the pristine (111) surface of $Bi_2Se_3$ are presented in Fig. 1(a). Each QL has the layer sequence Se-Bi-Se-Bi-Se. We considered two possible adsorption sites on the (111) surface of $Bi_2Se_3$ [A and B in Fig. 1(a)]. The A sites are the hollow sites of the nearest-neighbor $Se_3$ triangles on the surface Se layer, while the B sites are the on-top sites of the Bi atoms. Our calculations show that each Fe prefers the B-site adsorption to the A-site adsorption by 75.8 meV. Thus, hereafter, we will assume the B-site adsorption of Fe atoms unless otherwise stated. We denote the Bi and Se atoms adjacent to the Fe adatom at B site as Bi* and Se*, respectively. The Fe atom lies slightly (0.268 Å) above the center of the Se*$_3$ triangle, and the Bi* atom goes down by 0.753 Å from the position of the pristine 1QL-slab [Fig. 1(b)]. Without loss of generality, we take the local z-axis along the c-axis, and set the positive x-axis for a Fe atom pair along the dimer (i.e., $\mathbf{R} = \mathbf{r_2} - \mathbf{r_1}$).

*Electronic structure.* The band structure calculated for the pristine 1QL-slab [Fig. 2(a)] shows a band gap of 0.5 eV in agreement with experiment [18]. The lower Dirac cone disappears for the 1QL-slab [2,18], so only the upper Dirac cone is present in Fig. 2(a). The major components of this band are the $6p_z$ orbitals of the Bi atoms and the $4p_x/4p_y$ orbitals of the surface Se atoms. This surface band consists of the spin-up and spin-down bands that are degenerate. The band structure of the 1QL-slab with a single Fe adsorbed in a 4×4×1 supercell [Fig. 2(b)] shows a splitting of the spin-up and spin-down Dirac bands, which is due largely to their strong interactions with the spin-polarized Fe

$3d_{z^2}$ states. The Fe $3d_{z^2}$ interacts with the Bi* $6p_z$, and the Fe $3d_{xy}$ and $3d_{x^2-y^2}$ with the Se* $4p_x$ and $4p_y$. Consequently, the spin-polarized electrons of the Fe adatom interact strongly with the conducting electrons of the TSS. The calculated moment of the Fe adatom (~2.9 $\mu_B$) suggests the presence of approximately three unpaired electrons at the Fe site. The projected density of states (PDOS) calculated for the Fe 3d states [Fig. 2(c)] suggests an approximate crystal-field split pattern of the Fe 3d states depicted in Fig. 2(d). The approximate electron configuration of Fe is a high-spin in between $d^6$ and $d^7$.

*Apparent single-ion anisotropy.* In opening a gap at the Dirac point by magnetic adatoms, it is necessary that the adatoms have the //c-spin orientation [3]. To determine the preferred spin-orientation due to the SOC effect, we calculate the SIA energy, $A = E_{//c} - E_{\perp c}$, using one Fe adatom per 4×4 supercell, where $E_{//c}$ and $E_{\perp c}$ are the energies with adatom spin-orientation parallel and perpendicular to the c-axis, respectively. Our calculations show that A = -5.78 meV per Fe adatom for the B-site adsorption (A = -5.16 meV per Fe adatom for the A-site adsorption) when SOC effects are included on both the adatom Fe and the host $Bi_2Se_3$, showing that the Fe spin strongly prefers to align along the c-direction. Therefore, our finding is entirely consistent with the experimental observation of a massive Dirac fermion state in $Bi_2Se_3$ by chemical vapor deposition of Fe atoms on its (111) surface [3]. The above calculation of the SIA energy is for dilute adsorption coverage (6.25% for one Fe adatom per 4×4 supercell). Thus, the effect of the spin exchange on the SIA energy would be negligible (see below). For the B-site adsorption, the $E_{\perp c}$ exhibits small in-plane anisotropy (see Fig. S3 [20]). For simplicity, the average of the $E_{\perp c}$ values along the x- and y-axes will be used as $E_{\perp c}$ throughout this

Letter.

We now examine why a Fe adatom on $Bi_2Se_3$ (111) surface exhibits apparently a large SIA energy by considering the B-site adsorption. We already have A = -5.78 meV/Fe when SOC effects are included on both the adatom Fe and the host $Bi_2Se_3$. Further calculations show that A = -1.67 meV if SOC is allowed only on the Fe adatom, A = -4.93 meV if SOC is allowed only on the host $Bi_2Se_3$, and A = -4.91 meV if SOC is allowed only on the Fe, Bi* and Se* atoms. Thus, the apparent SIA of a Fe adatom arises largely from the Bi* and Se* atoms surrounding it. Since the spin-orbit interaction $\lambda L \cdot S$ is a local interaction [20,21], the above results indicate that the magnetic atom Fe induces spin polarization largely on the Bi* and Se* atoms, which gives rise to small spin moments on them, and the SOC associated with these small spin moments leads to the apparent SIA of the Fe adatom ([20], see below for further discussion). Thus, the apparent SIA of a Fe adatom, which is an important consequence of the adatom interaction with the TSS electrons, should be included in a model Hamiltonian describing the adatom interactions on TI surface because it has a significant influence on the spin orientation (see below, [20]).

*Interaction between adatoms*. To describe how two magnetic adatoms interact on the surface of a TI, we use the Hamiltonian [13,20,21]

$$H = \left(J_{zz}S_1^z S_2^z + J_{xx}S_1^x S_2^x + J_{yy}S_1^y S_2^y\right) + \mathbf{D} \cdot (\mathbf{S_1} \times \mathbf{S_2}) + A[(S_1^z)^2 + (S_2^z)^2], \qquad (3)$$

where the first term describes the Heisenberg-Ising spin exchange between the adatoms, the second term the DM interaction between them , and the last term is the SIA of the two adatoms, which should be included in Hamiltonian analysis according to our above discussions [20,21]. To determine the parameters $J_{ii}$ (i = x, y, z), **D** and A, we carry out energy-mapping analysis [13] on the basis of density functional calculations by treating the spin as unit vectors, i.e., $|\mathbf{S_i}| = 1$ (i = 1, 2) [20]. As shown in Fig. 1(a), we consider five different pairs of Fe adatoms at the B-sites and evaluate their $J_{ii}$ (i = x, y, z) values. For a pair of Fe adatoms occupying nearest-neighbor B sites, the relaxed distance R is 4.009 Å (hereafter R ≈ 4.0 Å), slightly shorter than the lattice constant 4.138 Å. In calculating the $J_{ii}$ values, we use the 4×4×1 supercell for the adatom pairs with R ≈ 4.0 and 7.2 Å, and the 8×4×1 supercell for those with R ≈ 8.3, 11.0 and 12.4 Å to avoid the direct interaction between adjacent pairs.

For the pair with R ≈ 4.0 Å, the spin exchanges are antiferromagnetic (AFM) and anisotropic, i.e., $J_{zz}$ = 32.9 meV, $J_{xx}$ = 34.9 meV, $J_{yy}$ = 27.1 meV. If the SOC on Fe is switched off, these spin exchanges hardly change. However, on switching off the SOC on the host $Bi_2Se_3$, the spin exchanges and their anisotropy become strongly reduced, namely, $J_{zz}$ = 18.3 meV, $J_{xx}$ = 18.8 meV, and $J_{yy}$ = 18.8 meV. Thus, the conducting electrons of the TSS enhance the spin exchange between the Fe adatoms. For the Fe adatom pair with R ≈ 7.2 Å, the spin exchanges are ferromagnetic (FM) and anisotropic, i.e., $J_{zz}$ = −9.8 meV, $J_{xx}$ = −10.0 meV, $J_{yy}$ = −15.3 meV. As summarized in Fig. 3 for the $J_{zz}$ values, as an example, the sign of the spin exchange oscillates as R increases, following a RKKY-behavior. This is in agreement with the conclusion from the model Hamiltonian studies [8,9,11]. The $J_{zz}$ values calculated without SOC on all atoms, also plotted in Fig. 3, show that the spin exchanges between adatoms are still strong even without SOC and follow a RKKY-behavior. Namely, under the mediation of the TSS electrons, the spin exchange is a long-range interaction [3, 11].

The DM vector **D** is also evaluated by performing energy-mapping analysis [13]. Symmetry analysis of the structures of the Fe adatom dimers shows that $D_x = 0$ for $R \approx 4.0$ Å, and $D_x = D_z = 0$ for $R \approx 7.2$ Å. The calculated DM vectors are consistent with this analysis; $D_x = 0$, $D_y = -10.1$ meV, and $D_z = 1.5$ meV for $R \approx 4.0$ Å, while $D_x = -0.1$ meV, $D_y = 3.8$ meV, $D_z = 0$ meV for $R \approx 7.2$ Å [20]. The nonzero component of the D vector is quite large compared with the spin exchange, i.e., $|D_y/J_{av}| \approx 0.3$, where $J_{av}$ is the average of $J_{xx}, J_{yy}$ and $J_{zz}$. To understand the origin of this large DM interaction, we selectively switch off the SOC on different atoms by considering the Fe dimer with $R \approx 4.0$ Å as an example. With SOC only on the $Bi_2Se_3$ host, $D_y = -10.8$ meV and $D_z = 0.5$ meV. With SOC only on the Fe adatoms, however, $D_y = -0.7$ meV and $D_z = 1.2$ meV. Thus, the DM interactions are also mediated by the conducting electrons of the TSS.

For the cases of two Fe adatoms on the (111) surface, the SIA A can also be evaluated by energy-mapping analysis [20], which shows that A depends on their distance R, e.g., A = -3.44 meV for $R \approx 4.0$ Å, and A = -6.63 meV for $R \approx 7.2$ Å. Nevertheless, the preference for the //c-spin orientation at the B-sites remains unaffected by the interactions between the adatoms mediated by the TSS electrons, nor by the thickness of the slab used for calculations [20].

*Effect of electric field*. Results of our calculations with electric field applied along the z-axis [20] are summarized in Fig. 4(a) for the case of the $R \approx 4.0$ Å Fe dimer. Increasing the electric field weakens the Heisenberg-Ising spin exchange but strengthens the DM interaction, raising the $|D_y/J_{av}|$ ratio up to ~0.6 in support of the finding by Zhu *et al.* [11]. The difference charge density plot obtained from the densities calculated for the electric fields of -1.2 V/Å and 0 [Fig. 4(b)] shows that, with negative electric field, electrons are transferred from the lower surface to the upper one, hence shifting the Fermi level and affecting the net magnetization as predicted by Zhu *et al.* [11], with the canted moment along the x-axis due to the large $D_y$. The field-induced change in the net magnetic moment is large (e.g., 0.54 $\mu_B$ at 0.4 V/Å and 1.18$\mu_B$ at -1.2 V/Å), suggesting the possibility of tuning the net magnetization by electric field [20].

Finally, we point out [20] that if the SIA is neglected, the spins of the magnetic adatoms should be nearly perpendicular to the c-axis and isotropic in the xz-plane. In contrast, they should be nearly parallel to the c-axis and anisotropic in all three directions when the SIA is not neglected. This difference can be experimentally tested.

*Conclusions.* The essential elements needed in describing the magnetic interactions between Fe adatoms on the surface of the TI $Bi_2Se_3$ are the anisotropic Heisenberg-Ising spin exchange, the DM interaction, and the SIA. All these effects originate substantially from the interactions of the magnetic adatoms with the TSS conducting electrons. The DM interaction is strong with $|D/J| \approx 0.3$, and the ratio can be further increased to ~0.6 by an electric field applied along the z-axis. The apparent SIA of magnetic adatoms should not be neglected.

*Acknowledgements.* This work is supported by NSF of China, the Special Funds for Major State Basic Research and the Research Program of Shanghai Municipality (Pujiang, Eastern Scholar).

*hxiang@fudan.edu.cn

†xggong@fudan.edu.cn

[1] H. J. Zhang et al., Nature Phys. **5**, 438 (2009).

[2] Y. Zhang *et. al.*, Nature Phys. **6**, 584 (2010).

[3] L. A. Wray *et. al.*, Nature Phys. **7**, 32 (2011).

[4] H. Chen et al., Phys. Rev. Lett. **107**, 056804 (2011).

[5] M. Z. Hasan and C. L. Kane, Rev. Mod. Phys. **82**, 3045 (2010).

[6] T. M. Schmidt, R. H. Miwa, and A. Fazzio, arXiv:1107.3810v1 (2011).

[7] J. Honolka *et. al*, arXiv:1112.4621v2 (2011).

[8] Q. Liu et al., Phys. Rev. Lett. **102**, 156603 (2009).

[9] R. R. Biswas and A. V. Balatsky, Phys. Rev. B **81**, 233405 (2010).

[10] C. W. Niu et al., App. Phys. Lett. **98**, 252502 (2011).

[11] J.-J. Zhu et al., Phys. Rev. Lett. **106**, 097201 (2011).

[12] R. Yu et al., Science **329**, 61 (2010).

[13] H. J. Xiang et al., Phys. Rev. B **83**, 174402 (2011).

[14] G. Kresse and J. Furthmüller, Phys. Rev. B **54**, 11 169 (1996).

[15] P. E. Blöchl, Phys. Rev. B **50**, 17 953 (1994).

[16] G. Kresse and D. Joubert, Phys. Rev. B **59**, 1758 (1999).

[17] J. P. Perdew, K. Burke, and M. Ernzerhof, Phys. Rev. Lett. **77**, 3865 (1996).

[18] Y. Sakamoto et al., Phys. Rev. B **81**, 165432 (2010)

[19] J. Neugebauer and M. Scheffler, Phys. Rev. B **46**, 16067(1992).

[20] See the Supplemental Material for details.

[21] D. Dai, H. J. Xiang, M.-H. Whangbo, J. Comput. Chem. **29**, 13 (2008).

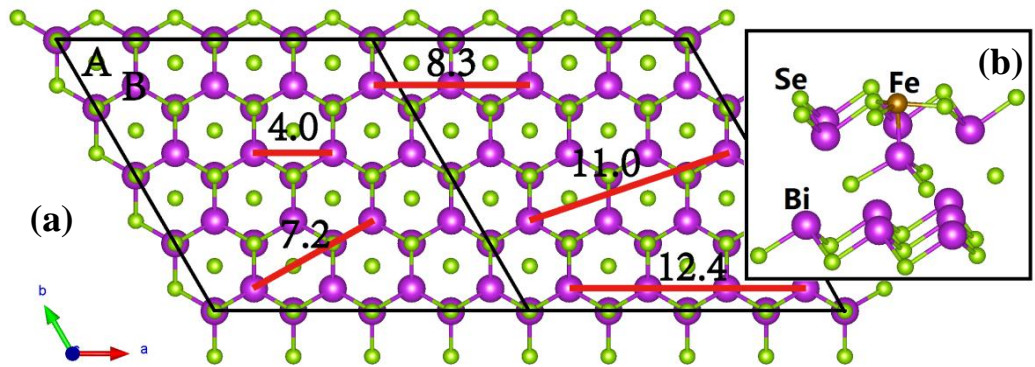

**Fig. 1** (color online). (a) A top view of the pristine (111) surface of $Bi_2Se_3$. A and B represent the two possible adsorption sites for a magnetic Fe atom. The positions for various pairs of Fe atoms at the B sites are indicated by red segments with their optimized distances in units of Å. (b) A side view of a 1QL-slab with a single isolated Fe adatom showing the distorted local structure around the Fe atom.

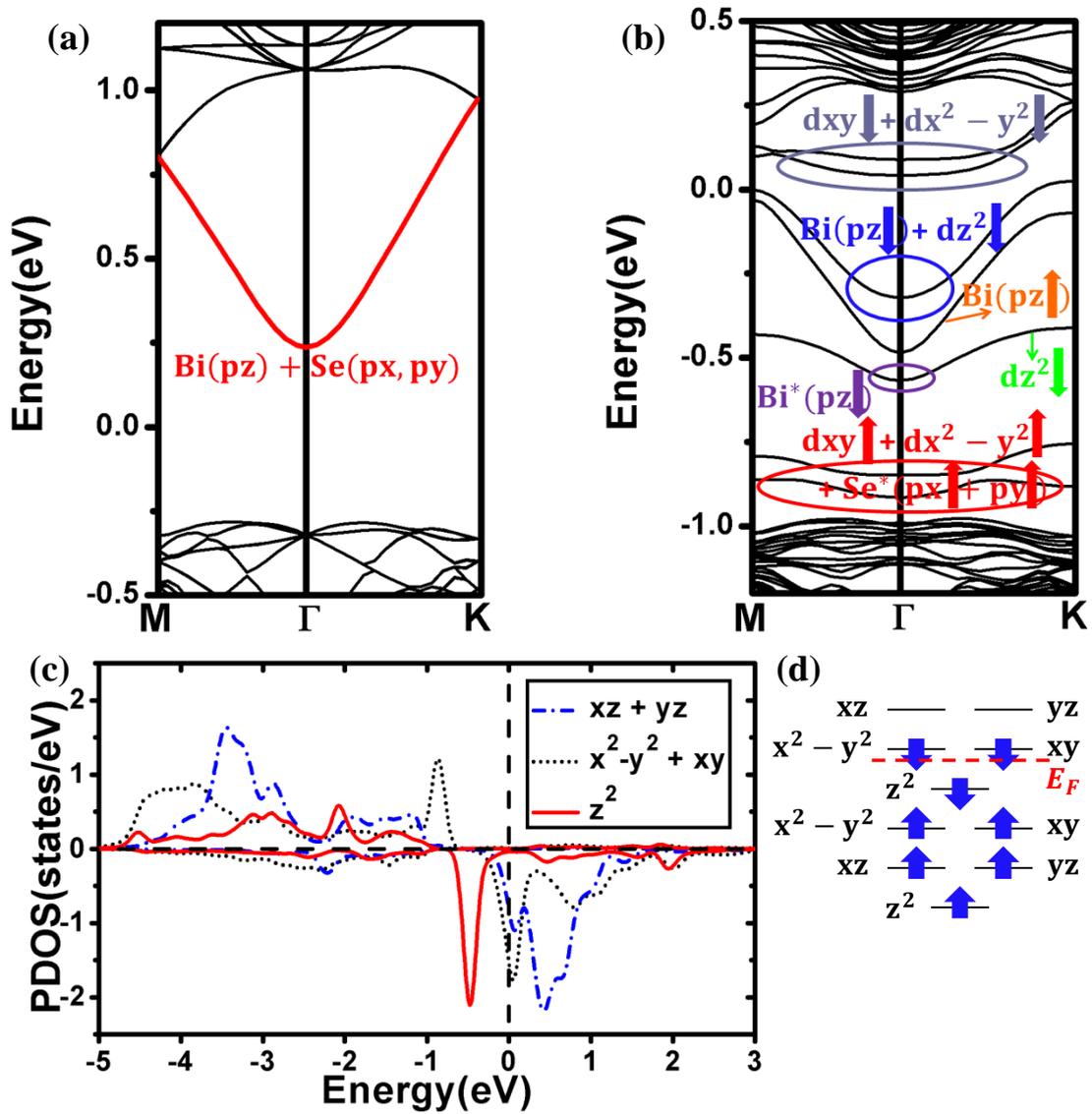

**Fig. 2** (color online). (a) The band structure calculated for the pristine 1QL-slab of $Bi_2Se_3$. The main surface components are marked in red. (b-d) The electronic structure of the 1QL-slab of $Bi_2Se_3$ with a single Fe atom adsorbed on the surface at B site. (b) The band structure showing that the Fe 3d states interact strongly with the 6p states of the Bi* atom and the 4p states of the Se* atoms. (c) The PDOS plots for the 3d states of the adsorbed Fe atom. (d) An approximate crystal-field split pattern for the 3d states of the adsorbed Fe atom and their electron occupation.

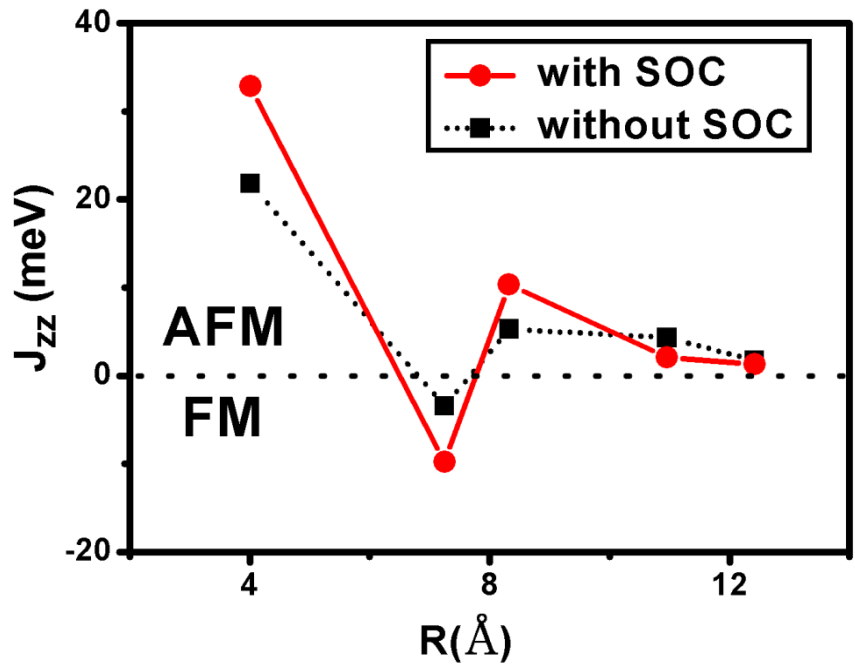

**Fig. 3** (color online). The dependence of the spin exchange $J_{zz}(R)$ between two Fe atoms adsorbed on the (111) surface on the distance R between them, calculated with and without SOC effects.

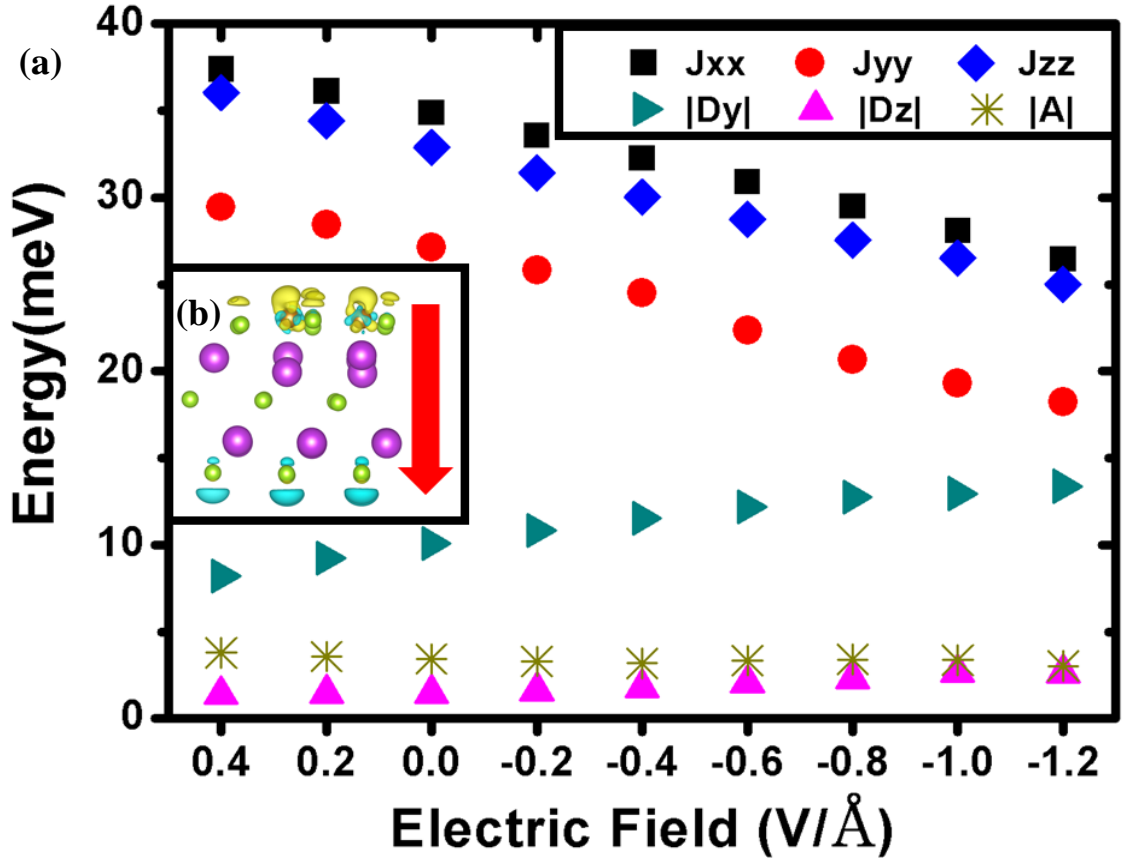

**Fig. 4** (color online). The effect of an electric field $E$ along the c-direction on the magnetic interaction parameters and the difference charge density distribution of the (111) surface of $Bi_2Se_3$ with Fe atoms adsorbed at the nearest-neighbor B sites: (a) The dependence of the spin exchange, the DM vector and the SIA term on $E$. (b) The difference charge density distribution, $\Delta\rho = \rho(E) - \rho(0)$, where $\rho(E)$ is the charge distribution under $E$ = -1.2 V/Å and $\rho(0)$ is that without applied electric field. The negative field points down in this figure as shown with the red arrow. The blue area denotes charge loss, and the yellow area charge accumulation.

Supplementary Material

for

**Strong single-ion anisotropy and anisotropic interactions of magnetic adatoms induced by topological surface states**

Z. L. Li, J. H. Yang, G. H. Chen, M.-H. Whangbo, H. J. Xiang, X. G. Gong

## 1. Apparent SIA of a magnetic adatom induced by the spin-polarization of the TSS

Our density functional calculations show that the magnetic adatom Fe spin-polarizes the electrons of the Bi* and Se* atoms. The small spin moments of Bi* and Se* induced by the spin polarization are antiparallel and parallel to that of Fe, respectively. To demonstrate that the polarization of the TSS, induced by Fe, leads to a strong apparent SIA, we consider the SOC effect on the p-states of Bi* as a representative example.

The PDOS plots calculated for the p-states of Bi* (Fig. S1) around the Fermi level reveal that the three p-states $|p_z,\uparrow>$, $|p_x,\downarrow>$ and $|p_y,\downarrow>$ of Bi* are nearly degenerate in the absence of SOC on $Bi_2Se_3$, but become split in the presence of SOC on $Bi_2Se_3$. We now examine this observation by considering how the $|p_z,\uparrow>$, $|p_x,\downarrow>$ and $|p_y,\downarrow>$ states become split under the action of SOC on Bi* using the SOC Hamiltonian $H_{SO} = \lambda L \cdot S$, where $\lambda$ is the SOC constant ($\lambda > 0$ for the case of one electron occupying the three p-states) and the $L \cdot S$ term is written as [S1]

$$L \cdot S = S_n \left(L_z\cos\theta + \frac{1}{2}L_+e^{-i\phi}\sin\theta + \frac{1}{2}L_-e^{i\phi}\sin\theta\right)$$
$$+ \frac{1}{2}S_+\left(-L_z\sin\theta - L_+e^{-i\phi}\sin^2\frac{\theta}{2} + L_-e^{i\phi}\cos^2\frac{\theta}{2}\right) \quad (S1)$$
$$+ \frac{1}{2}S_-(-L_z\sin\theta + L_+e^{-i\phi}\cos^2\frac{\theta}{2} - L_-e^{i\phi}\sin^2\frac{\theta}{2})$$

In terms of the spherical harmonics, the three p-states are expressed as

$$|p_z,\uparrow> = |Y_1^0,\uparrow>$$
$$|p_x,\downarrow> = |\frac{1}{\sqrt{2}}(Y_1^1 + Y_1^{-1}),\downarrow> \quad (S2)$$
$$|p_y,\downarrow> = |\frac{-i}{\sqrt{2}}(Y_1^1 - Y_1^{-1}),\downarrow>$$

We act the $H_{SO}$ on the three p-states to form the matrix M,

$$M = \begin{pmatrix} <p_z,\uparrow| \\ <p_x,\downarrow| \\ <p_y,\downarrow| \end{pmatrix} H_{SO} (|p_z,\uparrow> \quad |p_x,\downarrow> \quad |p_y,\downarrow>) \qquad (S3)$$

Due to the rotational symmetry of Bi*, we take $\phi = 0$ in Eq. (S1). We consider two spin orientations, $\theta = 0$ and $\frac{\pi}{2}$, for which the spin of Bi* is parallel and perpendicular to the z-axis, respectively.

(a) For $\theta = 0$, we have

$$M = \frac{\lambda \hbar^2}{2} \begin{pmatrix} 0 & 1 & -i \\ 1 & 0 & i \\ i & -i & 0 \end{pmatrix} \qquad (S4)$$

The eigenvalues and the corresponding eigenvectors are given as follow and in Fig. S2.

$$E_1 = -\lambda \hbar^2, \qquad \varphi_1 = \tfrac{1}{\sqrt{3}}(i|p_z,\uparrow> -i|p_x,\downarrow> +|p_y,\downarrow>)$$

$$E_2 = \tfrac{\lambda \hbar^2}{2}, \qquad \varphi_2 = \tfrac{1}{\sqrt{2}}(|p_z,\uparrow> +|p_x,\downarrow>) \qquad (S5)$$

$$E_3 = \tfrac{\lambda \hbar^2}{2}, \qquad \varphi_3 = \tfrac{1}{\sqrt{2}}(|p_z,\uparrow> +i|p_y,\downarrow>)$$

(b) For $\theta = \frac{\pi}{2}$, we have

$$M' = \frac{\lambda \hbar^2}{2} \begin{pmatrix} 0 & 0 & -i \\ 0 & 0 & 0 \\ i & 0 & 0 \end{pmatrix} \qquad (S6)$$

and

$$E'_1 = 0, \qquad \varphi'_1 = |p_x,\downarrow>$$

$$E'_2 = -\tfrac{\lambda \hbar^2}{2}, \qquad \varphi'_2 = \tfrac{1}{\sqrt{2}}(|p_z,\uparrow> - i|p_y,\downarrow>) \qquad (S7)$$

$$E'_3 = \tfrac{\lambda \hbar^2}{2}, \qquad \varphi'_3 = \tfrac{1}{\sqrt{2}}(|p_z,\uparrow> +i|p_y,\downarrow>)$$

As a result, the SOC favors the //z spin orientation ($\theta = 0$) for the case of one electron occupying the three states (Fig. S1). Note that the above discussion is valid for any fractional electron occupancy (less than 1) of the three levels, although we discussed as if there exists one whole electron to occupy them. The effect of SOC on the p-states of Se* can be discussed in a similar manner (not shown). As described in the main text, the contributions of the Fe, Bi* and Se* should be taken together as a whole in considering the SIA of a magnetic adatom Fe.

From our first-principles calculations, $A = E_{\parallel z} - E_{\perp z}$, which is consistent with the above conclusion that leading the Fe spin favors the //z orientation. We also find a slight in-plane anisotropy of $E_{\perp z}$ term, which could be depicted as an A vs. $\phi$ curve, as presented in Fig. S3. This in-plane anisotropy is substantial for the B-site adsorption of Fe, but negligible for the A-site adsorption. Nevertheless, the variation of this in-plane anisotropy (from -5.08 meV to -6.51 meV) for the B-site adsorption is small compared with the SIA, and we neglect it by taking the average of the maximum and the minimum SIA values, corresponding to ($E_{\parallel z} - E_{\parallel x}$) and ($E_{\parallel z} - E_{\parallel y}$), respectively.

## 2. Determination of the parameters for the spin Hamiltonian from first-principles density functional calculations by energy-mapping analysis

The spin Hamiltonian describing the interaction between Fe adatoms on the surface of a TI can be written as

$$H = \left(J_{zz}S_1^z S_2^z + J_{xx}S_1^x S_2^x + J_{yy}S_1^y S_2^y\right) + \mathbf{D} \cdot (\mathbf{S_1} \times \mathbf{S_2}) + A[(S_1^z)^2 + (S_2^z)^2]. \quad (S9)$$

To carry out energy-mapping analysis, we select different broken-symmetry spin configurations, and determine the energy of each state from first-principles calculations. We use the notation $E\left(S_1^{(S^x,S^y,S^z)}, S_2^{(S^x,S^y,S^z)}\right)$ to represent the energy of the specific spin configuration, where $S_i^{(S^x,S^y,S^z)}$ (i = 1, 2) is the spin orientation of each spin. Here we treat each spin as unit vectors, i.e., $|\mathbf{S_i}| = 1$ (i = 1, 2).

The Heisenberg-Ising terms are obtained as follows:

$$2J_{xx} = E\left(S_1^{(1,0,0)}, S_2^{(1,0,0)}\right) - E\left(S_1^{(1,0,0)}, S_2^{(-1,0,0)}\right),$$

$$2J_{yy} = E\left(S_1^{(0,1,0)}, S_2^{(0,1,0)}\right) - E\left(S_1^{(0,1,0)}, S_2^{(0,-1,0)}\right), \quad (S9)$$

$$2J_{zz} = E\left(S_1^{(0,0,1)}, S_2^{(0,0,1)}\right) - E\left(S_1^{(0,0,1)}, S_2^{(0,0,-1)}\right).$$

The direction of the DM vector $\mathbf{D}$ is constrained by the spatial structural symmetry [S1], and we first carry out the symmetry analysis and then perform the DFT calculations and check if they are consistent. In all cases, $\mathbf{R} = \mathbf{r}_2 - \mathbf{r}_1$ is set to be the positive direction of the x-axis. Considering the case of $R \approx 4.0\,\text{Å}$, this structure has a mirror plane perpendicular to $\mathbf{R}$, locating at the midpoint of the segment linking the two adatoms. In this case, $\mathbf{D}$ is parallel to the mirror plane, i.e., $D_x = 0$ [S1]. From the density functional point of view, we have

$$2D_x = E\left(S_1^{(0,1,0)}, S_2^{(0,0,1)}\right) - E\left(S_1^{(0,0,-1)}, S_2^{(0,-1,0)}\right),$$

$$2D_y = E\left(S_1^{(0,0,-1)}, S_2^{(-1,0,0)}\right) - E\left(S_1^{(1,0,0)}, S_2^{(0,0,1)}\right), \quad (S10)$$

$$2D_z = E\left(S_1^{(1,0,0)}, S_2^{(0,1,0)}\right) - E\left(S_1^{(0,-1,0)}, S_2^{(-1,0,0)}\right).$$

When $R \approx 7.2\,\text{Å}$, the structure has a mirror plane parallel to $\mathbf{R}$, containing the two adatoms. This would lead $\mathbf{D}$ perpendicular to the mirror plane, i.e., $D_x = 0$ and $D_z = 0$ [S1]. In terms of first-principles DFT calculations, we have

$$2D_x = E\left(S_1^{(0,1,0)}, S_2^{(0,0,1)}\right) - E\left(S_1^{(0,1,0)}, S_2^{(0,0,-1)}\right),$$

$$2D_y = E\left(S_1^{(1,0,0)}, S_2^{(0,0,-1)}\right) - E\left(S_1^{(1,0,0)}, S_2^{(0,0,1)}\right), \quad (S11)$$

$$2D_z = E\left(S_1^{(1,0,0)}, S_2^{(0,1,0)}\right) - E\left(S_1^{(1,0,0)}, S_2^{(0,-1,0)}\right).$$

The results from our DFT calculations are consistent with the symmetry analyses here.

The SIA energy A is simply given by $A = E_{\|z} - E_{\perp z}$ for the case of an isolated adatom. For the case of two adatoms interacting, it is obtained as

$$2A = E\left(S_1^{(0,0,1)}, S_2^{(0,0,1)}\right) - E\left(S_1^{(1,0,0)}, S_2^{(1,0,0)}\right) - J_{zz} + J_{xx},$$

or

$$2A = E\left(S_1^{(0,0,1)}, S_2^{(0,0,1)}\right) - E\left(S_1^{(0,1,0)}, S_2^{(0,1,0)}\right) - J_{zz} + J_{yy}. \quad (S12)$$

### 3. Effects of different Hamiltonian terms on the spin orientations

The Hamiltonian, Eq. (S8), consists of the Heisenberg-Ising term, the DM interaction term, and the SIA term. Here we discuss how these terms influences the orientations of the two Fe spins by using the parameters obtained for the case of $R \approx 4.0$ Å case, to show the SIA term is crucial for determining the spin orientation.

First, we consider only the Heisenberg-Ising exchange interactions. The positive $J_{ii}$ (i = x, y, z) makes the two spins have a collinear arrangement in the AFM state. If there is no anisotropy in the exchange interaction, the direction of these two spins is arbitrary. However, in our case, the anisotropy of $J_{zz}$, $J_{xx}$ and $J_{yy}$ would lead to a weak preference for the spins to point along the x-axis ($J_{xx} > J_{yy}, J_{zz}$).

Second, neglecting the spin exchange and the SIA, we consider only the DM interaction. The term $\mathbf{D} \cdot (\mathbf{S_1} \times \mathbf{S_2})$ forces the two spins to be perpendicular to each other to have the maximum energy gain, which gives rise to a competition with the spin exchange that prefers collinear spin arrangements. The large $D_y$ term makes the two spins perpendicular to each other in the xz-plane, and the small $D_z$ term makes the spins slightly out of the xz-plane.

Now we take the spin exchange and the DM interactions together. As a result of competition between the two interactions, we have AFM state along the x-axis dictated by $J_{xx}$ ($> J_{yy}$, $J_{zz}$). Because the moment cancels along the x-axis, there must be canted moments along the z- and the y-axes due to the DM interactions associated with $D_y$ and $D_z$, respectively [see Fig. S4(a)].

Finally, we take the effect of the SIA into account. Obviously, the large A forces the two spins to be (anti)parallel to the z-axis, overcoming their weak preference to align along the x-direction. Now, we have AFM state with spins along the z-axis, which utilizes the $J_{zz}$ component and has no conflict with A. In this case, the large $D_y$ forces the two spins to be aligned along the x-axis, thus leaving a canted moment along this direction. The $J_{yy}$ term puts the two spins antiparallel along the y-axis, and this is also favored by the small $D_z$ because a canted moment already exists along the x-axis [see Fig. S4(b)].

Taken together, the SIA makes the spins of the magnetic adatoms to align along the z-axis. Without the SIA, the spins lie nearly in the xy-plane because of the large $J_{xx}$ and $D_y$. In the model Hamiltonian study of Zhu *et al.* [S2], in which the SIA and $D_z$ terms are not included, and $J_{zz} = J_{xx} \neq J_{yy}$, there would be no anisotropy in the xz-plane. However, our calculations predict an anisotropic magnetic behavior along all the x-, y- and z- directions. This difference can be tested experimentally.

## 4. Simulation with 2QL- and 4QL-slabs

To validate the use of the results resulting from the 1QL-slab model of $Bi_2Se_3$, we calculate the $J_{zz}$, $J_{yy}$ and $D_y$ values for two Fe adatoms (with $R \approx 4.0$ Å) adsorbed on 2QL- and 4QL-slabs. It is noted that a pristine 4QL-slab exhibits a surface Dirac cone [S3], and a 3QL-slab was used to study the TSS by density functional calculations [S4]. We construct the structures of 2QL- and 4QL-slabs by using the 1QL-slab described as the topmost QL holding the Fe adatoms and then by adding other QL(s) taken from the pristine $Bi_2Se_3$ bulk below the topmost QL. Our calculations performed by using a 3×3×1 k-point set show that $A = -3.82$ meV, $J_{zz} = 36.75$ meV, $J_{yy} = 32.28$ meV and $D_y = -7.06$ meV for the 2QL-slab, while $A = -5.07$ meV, $J_{zz} = 33.79$ meV, $J_{yy} = 30.45$ meV and $D_y = -8.26$ meV for 4QL-slab. It is recalled that $A = -3.44$ meV, $J_{zz} = 32.88$ meV, $J_{yy} = 27.11$ meV and $D_y = -10.07$ meV for the 1QL-slab. Although the calculated values fluctuate slightly, the negative large A and the large |D/J| ratio remain unchanged, and the relationship between $J_{zz}$, $J_{yy}$ and $D_y$ in the 2QL- and 4QL-slabs is very similar to that observed for the 1QL-slab. These results ensure that our results from the calculations on the 1QL-slab are valid.

## 5. Effect of electric field

In their model Hamiltonian analysis of the RKKY interaction between magnetic atoms on the surface of a TI, Zhu *et al.* [S2] concluded that the Fermi level can be tuned by electric field, and the magnetization can be controlled as a consequence. Indeed, our calculations with electric field applied along the z-axis show that the magnetic interactions are delicately controlled by electric field. In our calculations the positive field direction is along the positive direction of the z-axis. Our 1QL-slab is too thin to apply a large electric field. However, with thicker slabs consisting of multiple QLs, a much stronger electric field can be applied to achieve $J_{ii} \approx 0$ so that a pure DM system can be realized, as suggested by Zhu *et al.* [S2].

With the parameters of the model Hamiltonian [Eq. (1)] determined by density functional calculations [Fig. 4(a)], one can derive the most stable spin configuration at a given electric field. Our results for the case of R ≈ 4 Å show that the total magnetic moment of the Fe pair cancels along the y- and z-axes (i.e., in AFM arrangement), but a tunable canted moment remains along the x-axis due to the large $D_y$. By analogy to the surface magnetoelectric effect observed for thin films [S5], we describe how the surface magnetic moment depends on applied electric field by calculating the surface magnetoelectric coefficient $α_s$

$$α_s = μ_0 ΔM/E, \qquad (S13)$$

where $ΔM$ is the magnetization introduced by the applied electric field $E$, and $μ_0$ is the magnetic permeability of free space. By linearly fitting the calculated magnetic moment as a function of $E$, we obtain $α_s \approx -2.8 \times 10^{-13} G cm^2/V$, which is much larger than the surface magnetoelectric coefficient of Fe film ($α_s \approx 2.9 \times 10^{-14} G cm^2/V$) [S5]. This suggests that the total magnetization of a TI possessing magnetic adatoms on its surface can be delicately controlled by electric field.

**References**

[S1] D. Dai, H. J. Xiang, M.-H. Whangbo, J. Comput. Chem. **29**, 13 (2008).

[S2] J.-J. Zhu, D.-X. Yao, S.-C. Zhang, and K. Chang, Phys. Rev. Lett. **106**, 097201 (2011).

[S3] T. M. Schmidt, R. H. Miwa, and A. Fazzio, arXiv:1107.3810v1 (2011).

[S4] H. Chen, W. G. Zhu, D. Xiao, and Z. Y. Zhang, Phys. Rev. Lett. **107**, 056804 (2011).

[S5] C.-G. Duan, J. P. Velev, R. F. Sabirianov, Z. Q. Zhu, J. H. Chu, S. S. Jaswal, and E. Y. Tsymbal, Phys. Rev. Lett **101**, 137201 (2008).

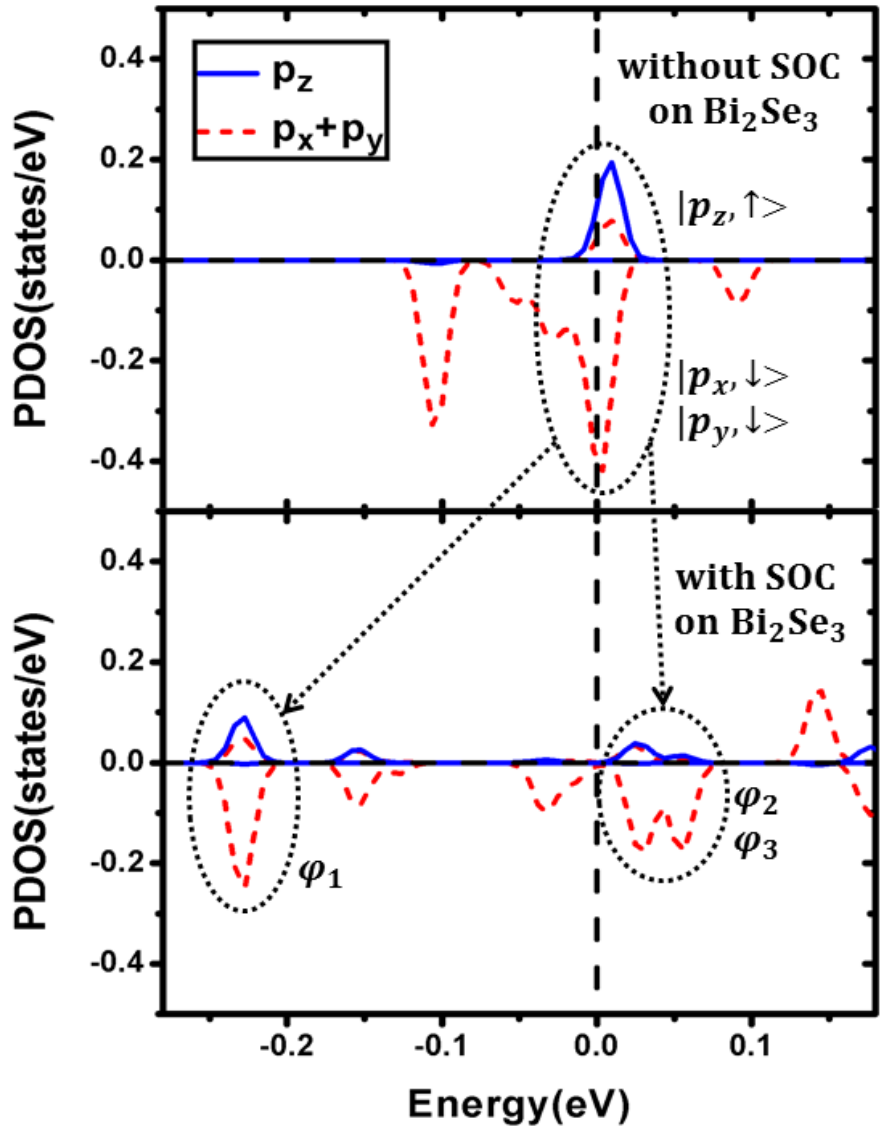

**Figure S1.** (color online). The PDOS plots calculated for the p-states of Bi* near the Fermi level before and after including effect of SOC. In the absence of SOC, the $|p_z,\uparrow\rangle$, $|p_x,\downarrow\rangle$ and $|p_y,\downarrow\rangle)$ states located around the Fermi level are practically degenerate. These states are split in energy when SOC is included.

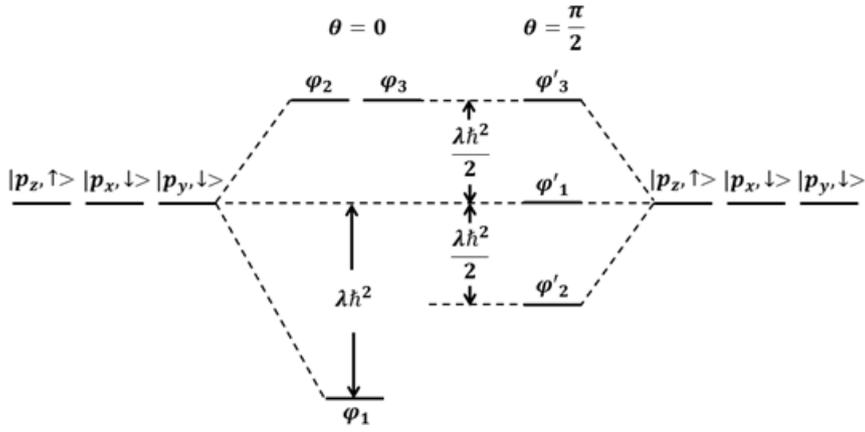

**Figure S2**. The split patterns of the three p-states of Bi* near the Fermi level (namely, $|p_z,\uparrow\rangle$, $|p_x,\downarrow\rangle$ and $|p_y,\downarrow\rangle$) under the effect of SOC when the spin orientation is along the z-axis ($\theta = 0°$) and perpendicular to the z-axis ($\theta = \frac{\pi}{2}$). With one electron to occupy the resulting three states, the SOC favors the z-axis orientation.

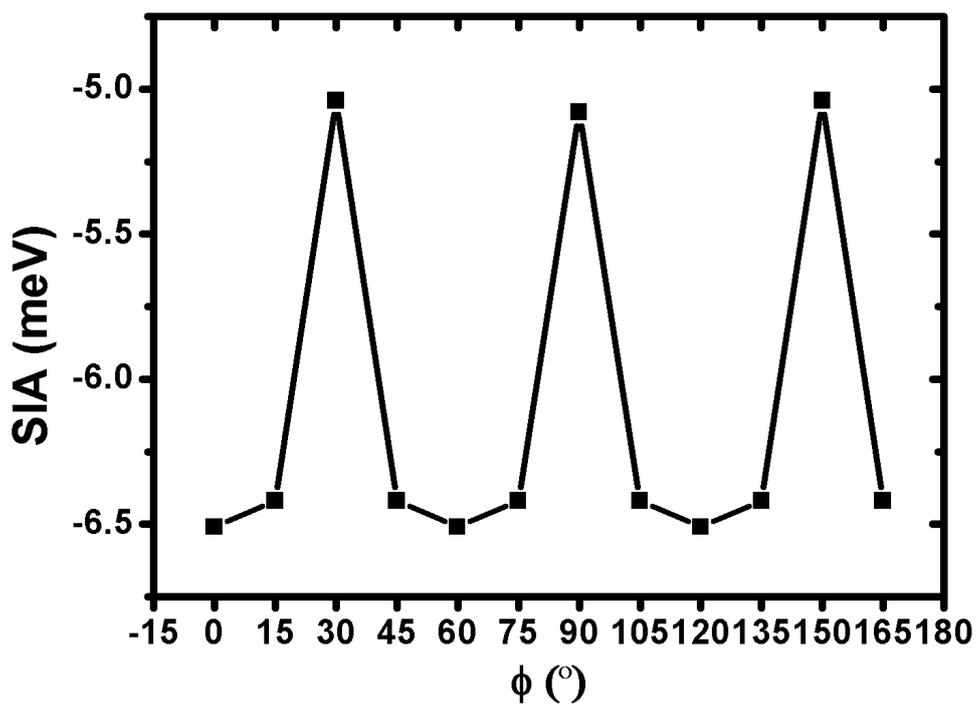

**Figure S3.** The SIA vs. ϕ curve calculated for the Fe adatom adsorbed on the B-site of Bi$_2$Se$_3$ (111), where the in-plane sweep angle ϕ is defined such that the direction for ϕ = 0° is parallel to one Se*-Se* edge of the Se*$_3$ triangle.

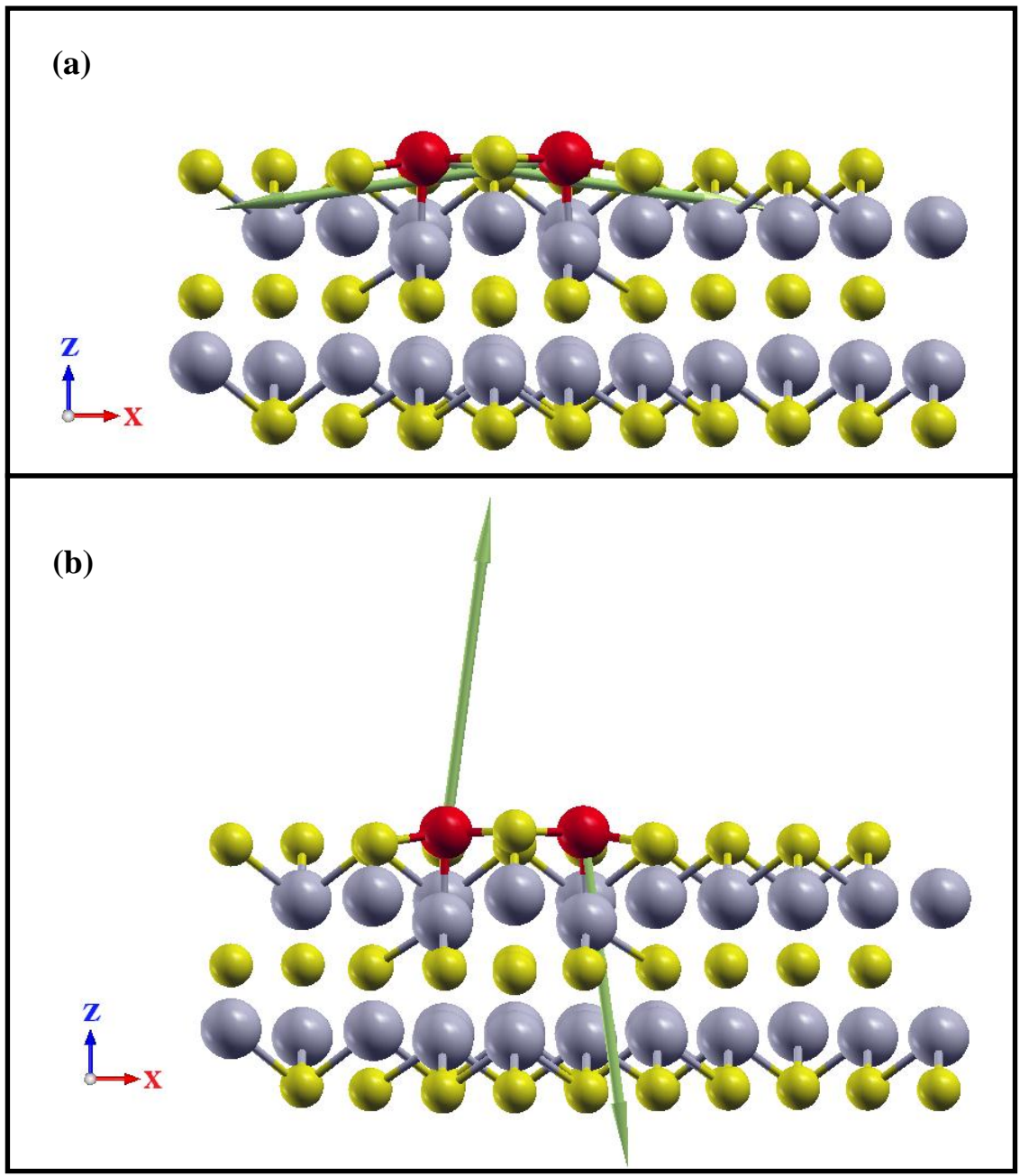

**Figure S4.** (color online). The spin orientations predicted (a) in the absence and (b) in the presence of MAE by using the model Hamiltonian with the parameters determined from first-principles DFT calculations.